# Critical analysis on efficiency and noise reduction methodologies of a switch mode power supply using LISN topology

L. Raveen S. De Silva[1]

[1] Department of Electrical Engineering, Faculty of Engineering, University of Moratuwa, Moratuwa, Sri Lanka

* Corresponding author: raveensdesilva@gmail.com

**Abstract**

This critical analysis offers a comprehensive overview of the efficiency and noise reduction methodologies applied in switch mode power supplies (SMPSs) and the emerging trends in the field. The study acknowledges that power supplies are an essential element in most electronic devices, and the recent developments have placed a greater emphasis on precision, miniaturization, and effectiveness of these components. The research focuses on SMPSs in comparison with other power supply methodologies and investigates the identification and mitigation of internal and external noise in DC Buck converter design through simulations. The study explores the latest and established technologies employed in improving SMPSs' power supply units and achieving maximum performance.

The paper highlights the use of a Linear Impedance Stabilization Network (LISN) topology and filter topologies to stabilize and eliminate noises in non-ideal scenarios, starting from an ideal buck converter. The research also emphasizes the design, analysis, and implementation of energy-efficient switch mode isolated converters commonly used in everyday life. Therefore, the study concludes that the integration of the LISN methodology in SMPSs design can play a crucial role in achieving maximum performance and efficiency.

## 1. Introduction

Computer simulation is a crucial tool in the design and analysis of components and systems across various engineering domains, particularly in power electronics converters and controllers. Designing power electronic systems without computer simulation can be labor-intensive, time-consuming, error-prone, and costly, with potential ramifications for component pricing [1,3]. Switched-mode power supplies (SMPSs) are a popular architecture for DC power supply in modern systems, ranging from simple mobile phone chargers to heavy industrial DC motors. Despite some limitations, the SMPS topology is widely recognized for its ability to achieve high efficiency, as well as spatial and thermal advantages that are critical considerations for power supply designers and engineers. To achieve a pure output from an SMPS, input and output noises originating from high-frequency switching and EMI transmission must be reduced [2-5].

This study focuses on the design of LISN model to reduce output noise from traditional SMPSs used with non-ideal components. A traditional DC/DC buck SMPS was employed, and filters were designed using LISN to suppress SMPS noise. Simulation results demonstrate that the designed filter significantly reduced SMPS output noise when compared to initial and final observations.

2. **Materials and methods**

*2.1 Analysis on noise interferences and current methods*

Switch mode power supplies (SMPS) are a powerful tool for designing electronic circuits due to their higher efficiency in energy conservation and ability to maintain a lower temperature in the circuit than linear voltage regulators. However, the increased complexity of SMPS requires extra effort during design and development to ensure proper operation. One critical aspect is the reduction of noise in the circuit to maintain its low amplitude, specifically voltage ripple and electromagnetic interference (EMI). This study primarily focuses on the reduction of output noises from a traditional SMPS using a filter for non-ideal components. Although a buck converter was used in this study, the methods outlined are applicable to various types of SMPS converters.

Voltage ripple is the small alternating current voltage present on top of the regulated direct current output of a SMPS. The reduction of this ripple is essential to ensure a clean DC supply and prevent digital circuitry failure and noise generation in analogue devices. The acceptable level of ripple amplitude varies depending on the device's sensitivity, but most devices operate within a voltage range of 50 to 100 microvolts. A datasheet check for the components and sufficient voltage gap between the power supply and the component's maximum voltage is essential when dealing with electrical components.

Electromagnetic interference (EMI) refers to unwanted electromagnetic emissions (EMEs) that interfere with other devices and wireless communication networks, which generate electrical impulses through the high-frequency switching of SMPS. SMPS generates EMI due to the output ripple noise, which may be transmitted or radiated, leading to interference in circuits that are otherwise functioning. The SMPS also generates electrical noise due to its switching frequency and the presence of reactive components. These noises disrupt the processes of the power supply, and the buck converter may broadcast electrical signals in the surrounding region, similar to a radio, under certain conditions.

Therefore, reducing the noise generated by SMPS is crucial. DC/DC converters are generally more efficient in voltage stabilization than linear regulators. However, they have a negative reputation in systems that contain sensitive signal route components due to the noise they generate, resulting in electromagnetic interference (EMI) emissions that can be transmitted or radiated. A simulation can be used to analyze the output and input switching EMI. The buck converter generates electrical noise due to its switching frequency and reactive components, leading to formation layout issues and potential electrical signal broadcasting in the surrounding region, similar to a radio. The circuit's numerous oscillations at a frequency higher than that of the MOSFET's switching frequency may interfere with the power supply's processes.

*2.2 Designing the general switch mode SMPS*

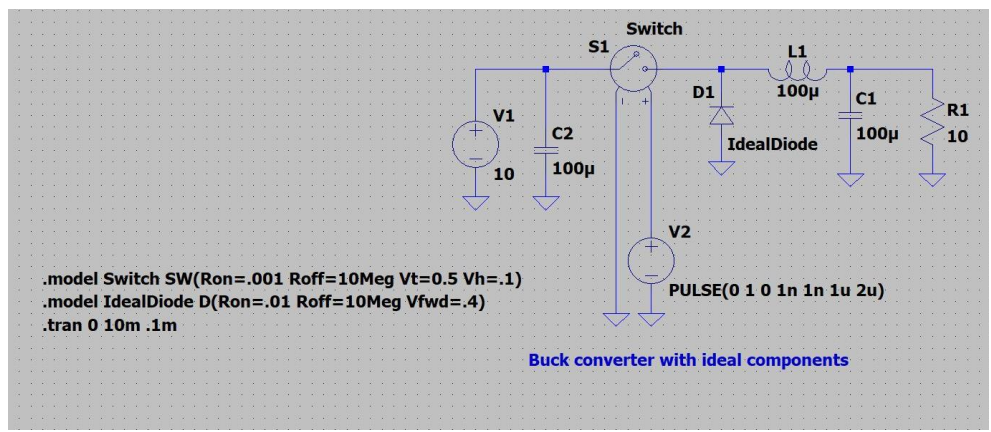

**Figure 1:** Switch mode DC/DC Buck converter with only essential components

A typical Switching Buck converter is composed of an integrated circuit (IC), switching transistors or MOSFETs, power stage components, a feedback network, and compensation and frequency settings. Due to the complexity of the system, simulation software may experience slower performance. In order to reduce the simulation time and focus solely on the noise generated by the power supply, a detailed analysis of the industrial SMPS feedback network is not necessary. Figure 1 demonstrates that for the purposes of noise analysis, only the essential power stage components, including the switch, diodes, inductor, input and output capacitors, and load, along with a driving signal, need to be simulated. Rather than aiming for highly accurate output voltages, the primary objective is to observe the switching behavior of the power supply, which is the main source of noise. To shorten the simulation time, efforts should be made to reduce the output stabilization period, which can result in a high level of oscillation during the initialization period, as seen in the first module.

*2.3 Improve the smps stabilizing time.*

Upon closer examination, it is apparent that the voltage was already nearing the desired output voltage of approximately 5 volts at the outset, and subsequently achieved this level with a short-lived oscillation. This suggests that similar measures may potentially be employed to expedite the power supply turn-on process. However, due to the oscillations in the power supply, all waveforms linked to it depict standard, unvarying input waveforms, which do not accurately reflect the outcomes of practical circuit simulations conducted in an expeditious manner. In real-world scenarios, the waveforms (as illustrated in Figure 2) do not appear as shown in the Buck converter ideal switching. At each of these edges, there are a plethora of differences, despite the appearance of uniformity during this phase.

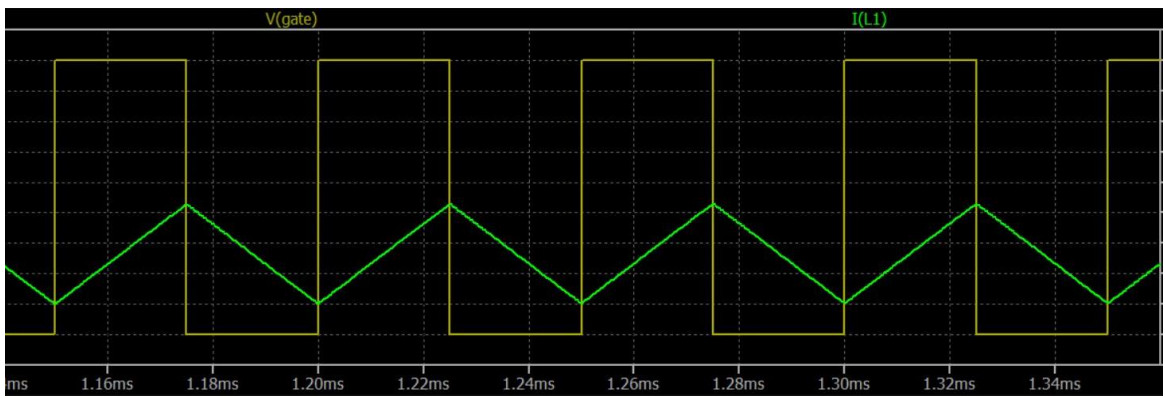

**Figure 2:** Buck converter Ideal switching

*2.4 Simulating the actual noises, and non-ideal conditions*

The next stage in replicating actual noise entails replacing some of the ideal components with real-world components. Upon scrutinizing the same waveforms, it becomes apparent that the current through the inductor remains nearly the same, but the voltage at the switch node begins to vary somewhat. This suggests that a real diode has replaced the ideal diode, and a P-MOS transistor has taken the place of the ideal switch. The output exhibits a specific fall time, a specific rise time, and numerous delays in between. As a result, the simulation starts to approximate the real component. Along with switching components, passive components are also taken into account. For instance, an inductor has a specific series resistance, which is invariably specified in the datasheet, as well as a certain parallel capacitance that can be estimated by utilizing the inductor's resonance frequency, which is also indicated in the datasheets.

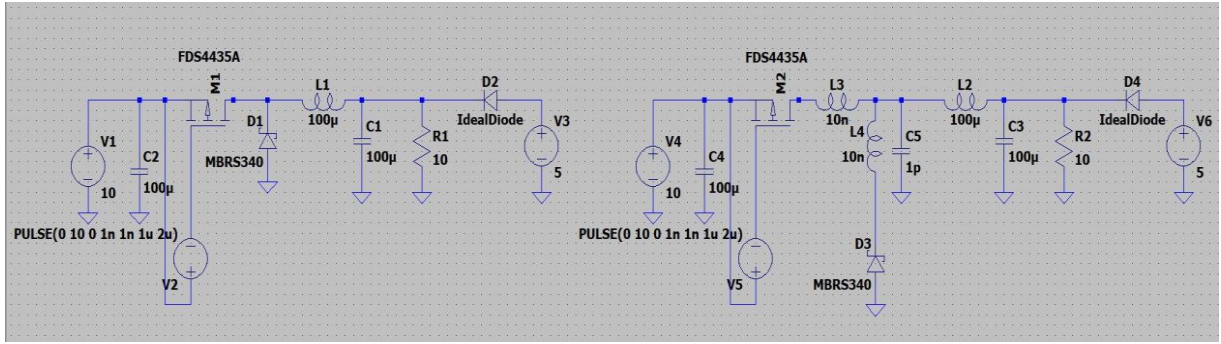

**Figure 3:** Schematic for Switch mode DC/DC Buck converter after adding Non idealities

The presence of interference in circuit architecture is a common phenomenon due to the non-ideal arrangement of components, resulting in impedance and mutual inductance effects that must be considered during the circuit's design phase. The impact of this interference on circuit performance must be investigated to ensure optimal functionality. During such an investigation, the presence of noise in the switching node was observed, causing the simulation to run at a slower pace. This occurrence is a typical phenomenon in schematics, and the level of noise is dependent on the performance of parasitic elements in the SMPS. To develop a realistic simulation, it is imperative to address this scenario before attempting to resolve the noise issue through simulation. The level of distortion added to the waveform is depicted in the figure, highlighting the primary objective of eliminating the effect on the output waveform.

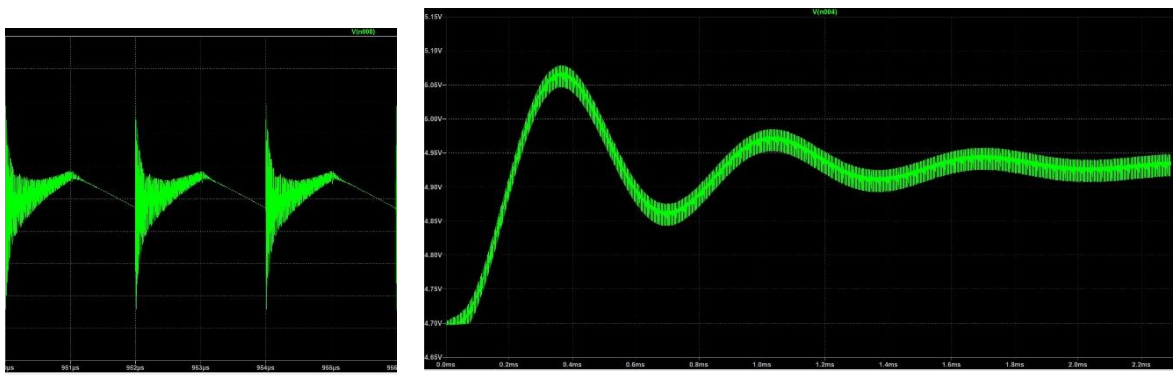

**Figure 4:** Alteration of the Switching Input

**Figure 5:** Noisy output of the SMPS before the steady state

The impact of added noise on the output waveform is vividly illustrated by Figures 4 and 5. The voltage ripple in the waveform exhibits a significant increase and only begins to stabilize around 4.95V over time, thereby indicating a notable level of instability. Furthermore, an elevated error percentage is conspicuously evident in the figures, underscoring the adverse impact of the added noise on the circuit's performance.

*2.5 Reduction of noises through an output filter*

Having obtained the circuit and exposed it to noise, the output exhibits a conspicuous level of noise, notably high-frequency noise, as evidenced by the simulation results. The occurrence of such noise is observed to take place predominantly at a frequency of approximately 100 MHz. To mitigate this noise, a practical solution would be to employ an inductor and capacitor filter. This approach is employed in the design depicted in Figure 6, which utilizes a 10uF capacitor and a 1uH inductor to counteract the noise.

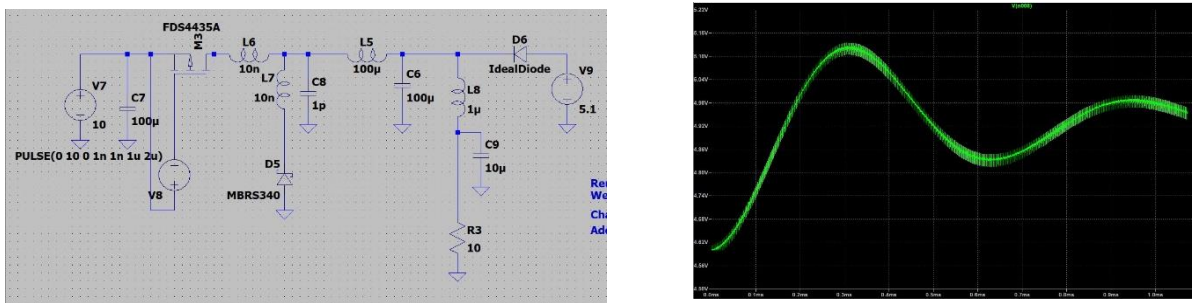

**Figure 6:** Adding a simple LC filter for the output terminal

**Figure 7:** Alteration of output waveform after Adding the LC filter

The aforementioned analysis reveals that the proposed method leads to a considerable reduction in high-frequency noise, as compared to Figure 7. Consequently, it is feasible to utilize an inductor with a low value that can demonstrate a high performance in contrast to its counterpart. Although the experimental setup included an ideal capacitor and an inductor with values of 100 MHz, the focus should be placed on the output capacitor. It is recommended to incorporate lower resistance components in the system to enhance energy efficiency. Additionally, a Buffered RC active filter is suggested, whereby the buffering function facilitates the maintenance of high efficiency while minimizing distortions in the output.

*2.6 Reduction of noises through a series RC filter*

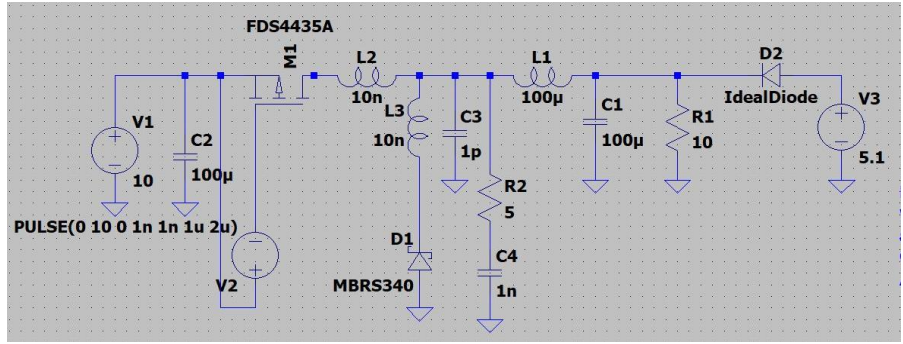

**Figure 8:** Schematic after adding a Series RC filter on the MOSFET

One potential approach for mitigating the noise is to implement filtering mechanisms, given that the noise was generated during the connection of inductors to the switching node. The use of snubbers, which involves the incorporation of series resistors, is a commonly employed and straightforward strategy within the industrial domain, as reported in literature. The implementation of snubbers has facilitated a significant reduction in simulation time, enabling the detection of previously undetectable noise sources. While the adoption of this filter effectively eliminates the majority of high-frequency noise, there is a notable power dissipation associated with the use of resistors, which represents a limitation of this approach. Despite this limitation, the snubber-based filtering approach is a simple and efficacious method for reducing background noise originating from source and mid schematic components.

*2.7 Countering non-ideality of the voltage input and adding LISN methodology*

Two feasible approaches to address the output noise and commonly occurring power supply noise are filtering the transistor command and introducing a delay in the rise and fall times of the transistor. Nonetheless, there exists an additional form of noise, known as internal noise, which has not been replicated within the simulation process. Furthermore, there is no observable input noise regardless of the nature or quality of the ideal voltage source employed, as evidenced by the absence of noise at all stages of voltage generation in the simulations conducted above. To rectify this imbalance, it is recommended to approximate the voltage signal as closely as possible, as exemplified in Figure 9a.

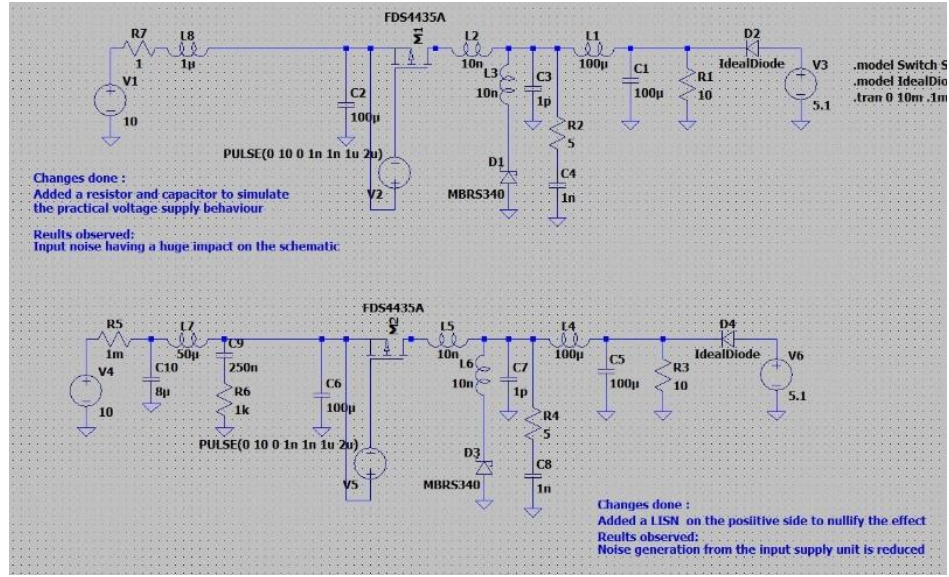

**Figure 9 A:** Schematic after adding alterations for the input Voltage source

**Figure 9 B:** Adding a LISN unit for the positive terminal

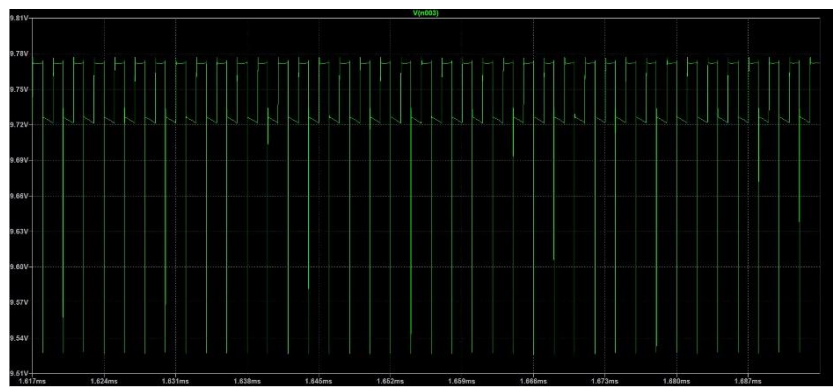

**Figure 10:** Input Voltage variation after adding non-ideality to the voltage supply

It is worth noting that connections between the power source and other components inevitably involve some level of resistance and inductance, as exemplified in Figure 13. This may result in the appearance of noise in the input signal, as evidenced by the input waveform shown in Figure 10. Addressing this issue requires the implementation of a Line Impedance Stabilization Network (LISN), which is a commonly utilized circuit schematic for emission testing purposes. The LISN methodology is particularly useful for compliance testing of various electronic units, including those employed in vehicle and component production. The LISN typically consists of a standard inductor and several capacitors connected to the power source on one side and the unit on the other side. The component values employed in the LISN network vary based on the simulated scenario. As illustrated in Figure 9b, incorporating a LISN network into the power supply circuit can effectively mitigate input noise and enable the extraction of a dedicated output centered around zero volts.

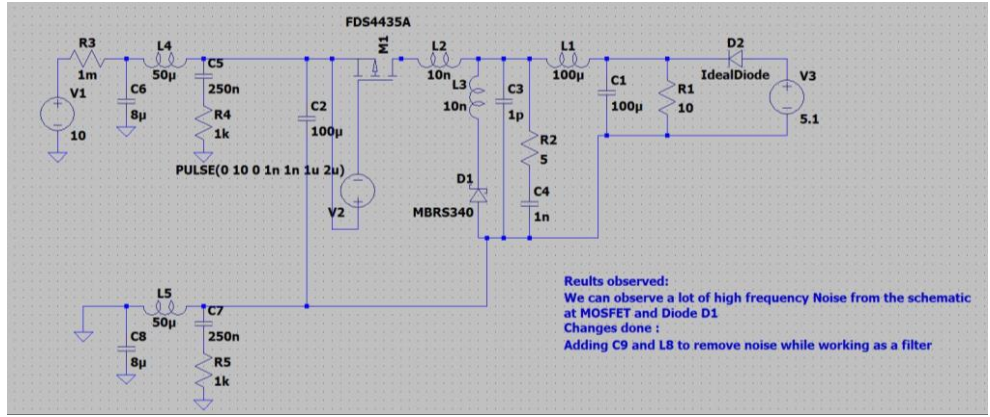

**Figure 11 : Adding LISN units for the positive and negative terminals**

When considering the addition of the LISN network, it is important to note two key points. Firstly, although it is currently only being used on the plus (+) supply, it should also be utilized on the minus (-) supply. This would entail connecting each input lead to the supply via the LISN network, which would prove highly convenient. Secondly, the capacity between the circuit and the ground has been omitted, thereby preventing the circuit ground from merely connecting to the earth through the LISN network, which is a crucial component. To ensure the most accurate results, it is essential to simulate the circuit as closely as possible to its real-world conditions, with particular attention given to replicating the parasitic capacitance. The accuracy of the conclusions drawn will be directly proportional to the precision with which this capacitance is replicated.

3. **Results and Discussion**

The study is brought to a close with the presentation of the final optimized schematic, illustrated in Figure 11. Throughout the phases of the study, various measures were implemented, including the incorporation of LISN topology, snubbers, and other strategies designed to mitigate the negative effects of noise distortions commonly associated with a conventional Switch mode Buck converter. The simulation results are displayed in the diagrams that follow.

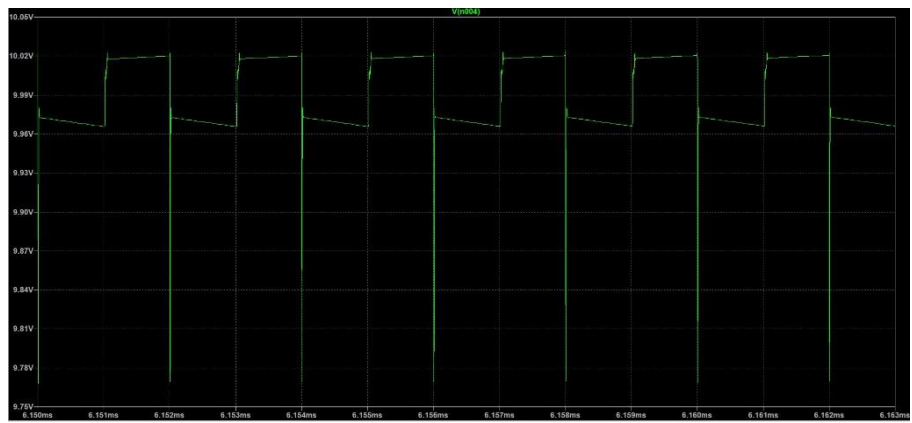

**Figure 12:** Variation of the Non-ideal Input voltage after adding LISN methodology and pi filter

In Figure 12, the time-dependent variation of the input DC value of 10 V sourced from a non-ideal power source, as it passes through the LISN networks, is presented, and can be compared to the input obtained during the first phases. By implementing the strategies described earlier, a more stable input waveform was obtained, as well as an improved output waveform. Furthermore, as demonstrated in Figure 13, the distribution of the waveform and its tolerance have been significantly reduced.

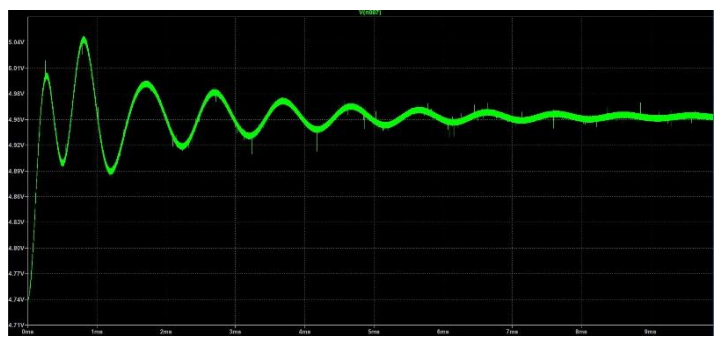

**Figure 13:** Variation of the Output voltage after adding LISN methodology and the pi filter

In comparison to the findings from the first phases, Figures 13 exhibit a considerable increase in the tolerance and rippling time of the waveform, implying a notable shift in the graph's behavior. It is evident that both issues were successfully addressed, as reflected in Figures 14A and 14B, which present the study's concluding FFT outputs. The results demonstrate a considerable reduction in EMI noise achieved through the implementation of numerous strategies, which were executed throughout the various phases of the study and presented in the results section.

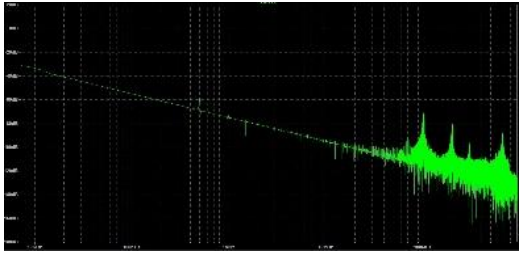 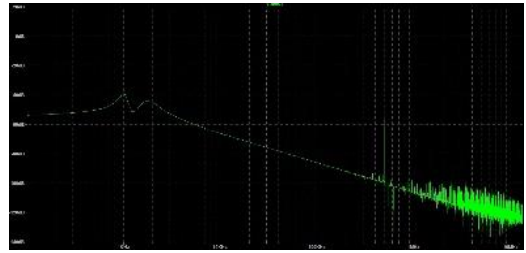

**Figure 14 A:** FFT analysis without adding the LISN topology

**Figure 14 B:** FFT analysis after adding LISN topology

### 4. Conclusion

The ever-increasing switching frequencies for SMPS necessitate the reduction of scale, budget, and manufacturing timelines within the industry in order to facilitate business growth. This has placed immense pressure on engineers to rapidly improve their functional designs. The implementation of noise reduction measures requires the development of many variables, which may prove challenging. The use of a switch-mode power supply is common when sensitive loads are being supplied, and the electrical engineering profession recognizes the need for additional filtering to reduce noise in the voltage signals, which may cause non-ideal anomalies. In contrast to sensitive loads regulated by a buck converter, noise in conventional circuits may be connected in various ways, and distinguishing between the different types of noise can be difficult. The circuit board heats up significantly due to the alternating currents generated at the back stage's inlet, as the currents are routed over long distances regularly. Fluctuating current levels may be inductively coupled to other circuit components, leading to interference with sensitive load circuits. When simulating the noise that SMPS generate, it is essential to model various intricate integrated circuits that have no effect on the noise generated; otherwise, it would be a waste of time and effort to simulate them. The simulation period needs to be as brief as possible to allow for the greatest number of viable runs. The study conducted in the first phases endeavored to keep the simulations running as quickly as possible. In subsequent phases, actual components replaced ideal ones, and each anomaly was simulated separately to determine its effect on output distortions. The results of the simulations were compared to those obtained using ideal components, and the optimization of the switch mode buck converter was completed.

The use of LISN topology proved to be a critical noise reduction methodology in the study, particularly in the presence of sensitive loads. The incorporation of LISN networks into the circuit design facilitated more stable input and output waveforms, reduced waveform distribution and tolerance, and minimized EMI noise. The results of the study demonstrate that the use of LISN topology in switch mode power supplies is an effective method of reducing noise and improving efficiency. Thus, the use of LISN topology should be considered an essential component of SMPS design, particularly when dealing with sensitive loads. Further research in this area could focus on exploring the use of advanced materials and manufacturing techniques to improve the performance of LISN networks and enhance their noise reduction capabilities.

5. Acknowledgment

It is important to acknowledge that the success of a project is not solely attributed to the efforts of the researcher, but also to the support and guidance provided by others. Therefore, I would like to express my gratitude to those who have played an instrumental role in the accomplishment of the project's objectives. I would like to extend my heartfelt appreciation to the individuals who have provided me with their invaluable support and assistance throughout the course of this study.